\documentclass [preprint,aps,showpacs] {revtex4}
\usepackage{bm}
\usepackage{epsfig}
\begin{document}

\title{Size quantization of charge carriers in lead salt cylindrical
quantum wires}

\author{S.V.~Goupalov}
\affiliation{Department of Physics, Jackson State University,
Jackson, MS 39217 USA\\
and A.F.~Ioffe Physico-Technical Institute, 26 Polytechnicheskaya, 194021 St.~Petersburg, Russia}

\begin{abstract}
A formalism for determining energy eigenstates of cylindrical
lead salt quantum wires in the multiple-band envelope-function approximation
is developed. Electron energy dispersion for quantum wire subbands within
the conduction and valence
bands is found.
\end{abstract}

\pacs{73.21.Hb,73.22.Dj,78.67.Lt}
\maketitle

The effect of quantum confinement on electron and hole states in
spherical quantum dots and cylindrical quantum wires of III-V and II-VI
semiconductor compounds having a complex valence-band structure was
described within the multiband envelope-function approximation more than
20 years ago~\cite{sercel}. Description of electronic
structure of spherical IV-VI semiconductor quantum dots
using similar approximation followed soon~\cite{kang}.
Recently nanostructures
based on IV-VI semiconductor compounds such as lead salts
received much attention due to their potential
for applications in solar cells and as infra-red detectors.
Quasi-zero-dimensional nanostructures of lead salts have become the
subject of wave-function engineering~\cite{bartnik,grinbom}, and there
appeared an interest in quasi-one-dimensional nanostructures of IV-VI
semiconductors~\cite{rupasov}. However, one is forced to accept
the fact that a recent attempt~\cite{rupasov} to
describe electronic structure
of a cylindrical lead salt quantum wire within a multiband
envelope-function formalism is not quite correct. This Brief Report seeks
to compensate for this deficiency.

The conduction and valence band extrema in lead salt semiconductors
(PbSe, PbS) occur at the $L$-points of the Brillouin zone.
Electron spectrum near the $L$-point
taking into account only the two closely lying conduction
and valence bands and neglecting band anisotropy can be described by the
spherical Dimmock model~\cite{kang,apokrif,falkovsky,myPbSe}.
In this model the electron
wave function is written as
\begin{equation}
\label{Psi1}
\Psi= \hat{u} \, |L_6^- \rangle + \hat{v} \, |L_6^+ \rangle \,,
\end{equation}
where $|L_6^- \rangle$ and $|L_6^+ \rangle$ describe
the Bloch functions while $\hat{u}({\bf r})$ and $\hat{v}({\bf r})$
are the spinors slowly varying with coordinates and satisfying the
equations~\cite{fn1}
\begin{equation}
\label{dimmock}
\left[
\matrix{
\left( \frac{E_g}{2} - \alpha_c \, \Delta \right)
&
-i P \left( {\bm \sigma} {\bm \nabla} \right)
\cr
-i P \left( {\bm \sigma} {\bm \nabla} \right)
&
-\left( \frac{E_g}{2} - \alpha_v \, \Delta \right)
\cr
}
\right] \,
\left[
\matrix{
\hat{u} \cr
\hat{v}
}
\right]
=E \, \left[
\matrix{
\hat{u} \cr
\hat{v}
}
\right] \,.
\end{equation}
Here $\sigma_{\beta}$ ($\beta=x,y,z$) are the Pauli matrices, $\alpha_c$,
$\alpha_v$, $E_g$, and $P$ are parameters of the model and $E$ is the electron
energy.

We will first construct linearly independent solutions of
Eqs.~(\ref{dimmock})
having cylindrical
symmetry and describing electronic states in a bulk semiconductor.
These solutions can be characterized by the energy, $E$, momentum, $k_z$
along the quantum wire axis, and projection, $M$ of the total angular
momentum on the wire axis.
Then we
will impose the boundary condition of the four-component envelope wave
function vanishing on the cylindrical surface of the quantum wire. In the
cylindrical coordinates $(\rho,\varphi,z)$ this boundary condition is to be
imposed at $\rho=R$, where $R$ is the radius of the cylindrical quantum wire.

The high symmetry of the problem allows us to separate variables and
pin down a set of good quantum numbers ($E$, $k_z$, and $M$).
After this Eqs.~(\ref{dimmock}) can be considered as a system of four
coupled ordinary linear differential equations
of second order with respect to the variable $\rho$.
As we are only interested in
solutions of Eqs.~(\ref{dimmock}) finite at $\rho=0$,
there are at most four such linearly independent
solutions of Eqs.~(\ref{dimmock}).

Let us look for a solution of Eqs.~(\ref{dimmock}) in the form
\begin{equation}
\label{ansatz1u}
\hat{u}_{M} (\rho,\varphi,z)=e^{i \, k_z \, z}
\left[
\matrix{
A \, e^{i \, (M-1/2) \, \varphi} \, J_{M-1/2}(k \rho) \cr
B \, e^{i \, (M+1/2) \, \varphi} \, J_{M+1/2}(k \rho) \cr
}
\right] \,,
\end{equation}
\begin{equation}
\label{ansatz1v}
\hat{v}_{M} (\rho,\varphi,z)=e^{i \, k_z \, z}
\left[
\matrix{
C \, e^{i \, (M-1/2) \, \varphi} \, J_{M-1/2}(k \rho) \cr
D \, e^{i \, (M+1/2) \, \varphi} \, J_{M+1/2}(k \rho) \cr
}
\right] \,,
\end{equation}
where $J_n(x)$ is the Bessel function of order $n$, $k$ is the wave number of transverse
motion, and $A$, $B$, $C$, and $D$ are
the coefficients to be determined.
Substitution of Eqs.~(\ref{ansatz1u}),~(\ref{ansatz1v}) into
Eqs.~(\ref{dimmock}) leads to a system of the following algebraic
equations:
\begin{equation}
\label{sys1}
P \, k_z \, A -i \, P \, k \, B - (\alpha_v \, k^2 + \alpha_v \, k_z^2
+E+E_g/2) \, C=0 \,,
\end{equation}
\begin{equation}
\label{sys2}
i \, P \, k \, A - P \, k_z \, B - (\alpha_v \, k^2 + \alpha_v \, k_z^2
+E+E_g/2) \, D=0 \,,
\end{equation}
\begin{equation}
(\alpha_c \, k^2 + \alpha_c \, k_z^2 -E+E_g/2) \, A+
P \, k_z \, C -i \, P \, k \, D=0 \,,
\end{equation}
\begin{equation}
(\alpha_c \, k^2 + \alpha_c \, k_z^2 -E+E_g/2) \, B+
i \, P \, k \, C - P \, k_z \, D=0 \,.
\end{equation}
The condition that this system of algebraic equations has a non-trivial
solution yields
\begin{equation}
\label{k}
k^2+k_z^2=\Xi+\Lambda \,,
\end{equation}
where
\[
\Lambda=\frac{E \, (\alpha_v-\alpha_c) - P^2 -E_g \, (\alpha_v+\alpha_c)/2}{2 \alpha_c \alpha_v} \,,
\]
\[
\Xi=\frac{\sqrt{\left[ E (\alpha_v- \alpha_c)-E_g (\alpha_v+\alpha_c)/2-P^2 \right]^2 +\alpha_c \alpha_v (4 E^2 - E_g^2)}}{2 \alpha_c \alpha_v} \,.
\]
Equations~(\ref{sys1}),~(\ref{sys2}) allow one to express the coefficients
$C$ and $D$ in terms of $A$ and $B$:
\[
C=\frac{P(k_z \, A-i \, k \, B)}{\alpha_v \, k^2 + \alpha_v \, k_z^2+E+E_g/2} \,,
\]
\[
D=\frac{P(i \, k \, A- k_z \, B)}{\alpha_v \, k^2 + \alpha_v \, k_z^2+E+E_g/2} \,.
\]
We see that a natural choice of the two linearly independent solutions will be
to set either $A$ or $B$ equal to zero. The resulting solutions (which turn
out to be orthogonal) take the form
\begin{equation}
\label{sol1u}
\hat{u}_{M}^{(1)} (\rho,\varphi,z)=A \, e^{i \, k_z \, z}
\left[
\matrix{
e^{i \,(M-1/2) \, \varphi} \, J_{M-1/2}(k \rho) \cr
0 \cr
}
\right] \,,
\end{equation}
\begin{equation}
\label{sol1v}
\hat{v}_{M}^{(1)} (\rho,\varphi,z)=\frac{i \, P \, A \, e^{i \, k_z \, z}}
{\alpha_v \, k^2 + \alpha_v \, k_z^2+E+E_g/2}
\,
\left[
\matrix{
-i \, k_z \, e^{i \, (M-1/2) \, \varphi} \, J_{M-1/2}(k \rho) \cr
k \, e^{i \, (M+1/2) \, \varphi} \, J_{M+1/2}(k \rho) \cr
}
\right] \,,
\end{equation}
\begin{equation}
\label{sol2u}
\hat{u}_{M}^{(2)} (\rho,\varphi,z)=B \, e^{i \, k_z \, z}
\left[
\matrix{
0 \cr
e^{i \, (M+1/2) \, \varphi} \, J_{M+1/2}(k \rho) \cr
}
\right] \,,
\end{equation}
\begin{equation}
\label{sol2v}
\hat{v}_{M}^{(2)} (\rho,\varphi,z)=\frac{i \, P \, B \, e^{i \, k_z \, z}}
{\alpha_v \, k^2 + \alpha_v \, k_z^2+E+E_g/2}
\,
\left[
\matrix{
- k \, e^{i \, (M-1/2) \, \varphi} \, J_{M-1/2}(k \rho) \cr
i \, k_z \, e^{i \, (M+1/2) \, \varphi} \, J_{M+1/2}(k \rho) \cr
}
\right] \,.
\end{equation}
The two remaining solutions of Eqs.~(\ref{dimmock}) can be sought in the form
\begin{equation}
\label{ansatz2u}
\hat{u}_{M} (\rho,\varphi,z)=e^{i \, k_z \, z}
\left[
\matrix{
C \, e^{i \, (M-1/2) \, \varphi} \, I_{M-1/2}(\kappa \rho) \cr
D \, e^{i \, (M+1/2) \, \varphi} \, I_{M+1/2}(\kappa \rho) \cr
}
\right] \,,
\end{equation}
\begin{equation}
\label{ansatz2v}
\hat{v}_{M} (\rho,\varphi,z)=e^{i \, k_z \, z}
\left[
\matrix{
F \, e^{i \, (M-1/2) \, \varphi} \, I_{M-1/2}(\kappa \rho) \cr
G \, e^{i \, (M+1/2) \, \varphi} \, I_{M+1/2}(\kappa \rho) \cr
}
\right] \,,
\end{equation}
where $I_n(x)$ is the modified Bessel function of order $n$ while $C$, $D$,
$F$, and $G$ represent a new set of the coefficients to be determined.
Substitution of Eqs.~(\ref{ansatz2u}),~(\ref{ansatz2v}) into
Eqs.~(\ref{dimmock}) yields
\begin{equation}
\label{sys3}
P \, k_z \, C -i \, P \, \kappa \, D + (\alpha_v \, \kappa^2
- \alpha_v \, k_z^2
-E-E_g/2) \, F=0 \,,
\end{equation}
\begin{equation}
\label{sys4}
-i \, P \, \kappa \, C - P \, k_z \, D + (\alpha_v \, \kappa^2 - \alpha_v \, k_z^2
-E-E_g/2) \, G=0 \,,
\end{equation}
\begin{equation}
(\alpha_c \, k_z^2 - \alpha_c \, \kappa^2 -E+E_g/2) \, C+
P \, k_z \, F -i \, P \, \kappa \, G=0 \,,
\end{equation}
\begin{equation}
(\alpha_c \, k_z^2 - \alpha_c \, \kappa^2 -E+E_g/2) \, D-
i\, P \, \kappa \, F - P \, k_z \, G=0 \,.
\end{equation}

\begin{figure}[ht]
  \centering
    \includegraphics[width=.6\textwidth]{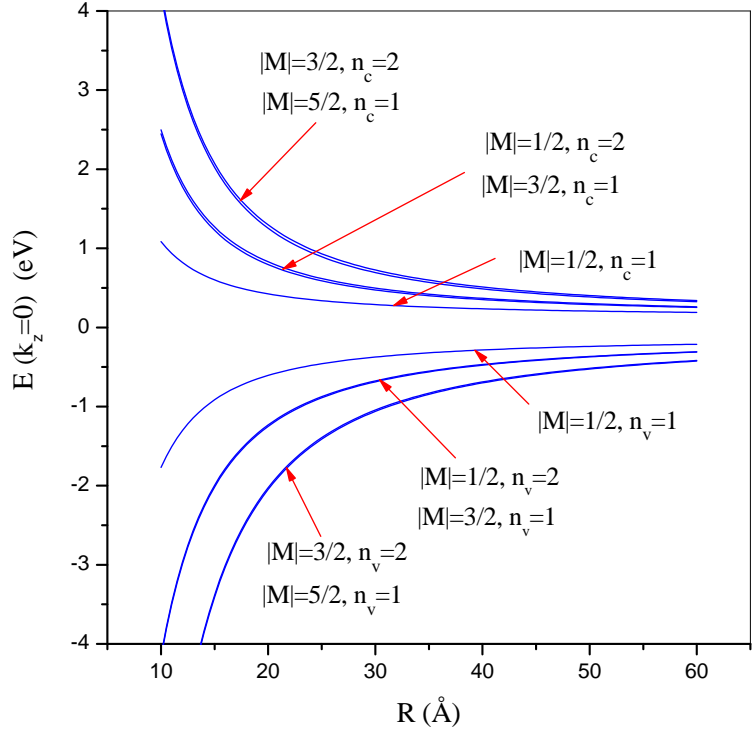}
\caption{(Color online) Energy of electron states in PbSe
cylindrical quantum wire at $k_z=0$
as a function of the wire radius.}
\end{figure}

\begin{figure}[ht]
  \centering
   \includegraphics[width=.6\textwidth]{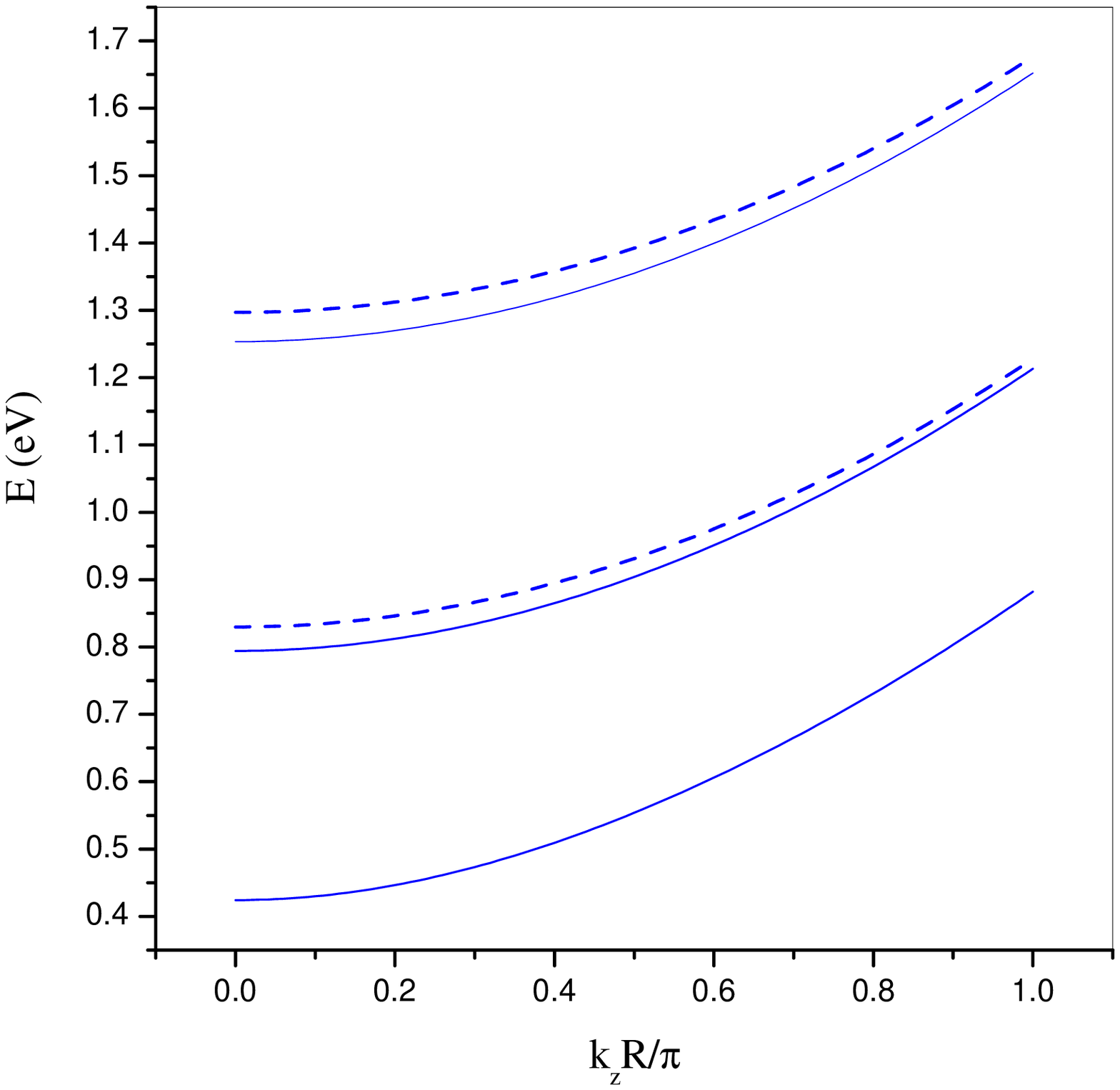}
\caption{(Color online) Energy dispersion of electron subbands
within the conduction band of a PbSe
cylindrical quantum wire of radius $R=20$~\AA.
By dashed lines are shown subbands
with the quantum number $n_c=2$.}
\end{figure}

\begin{figure}[ht]
  \centering
   \includegraphics[width=.6\textwidth]{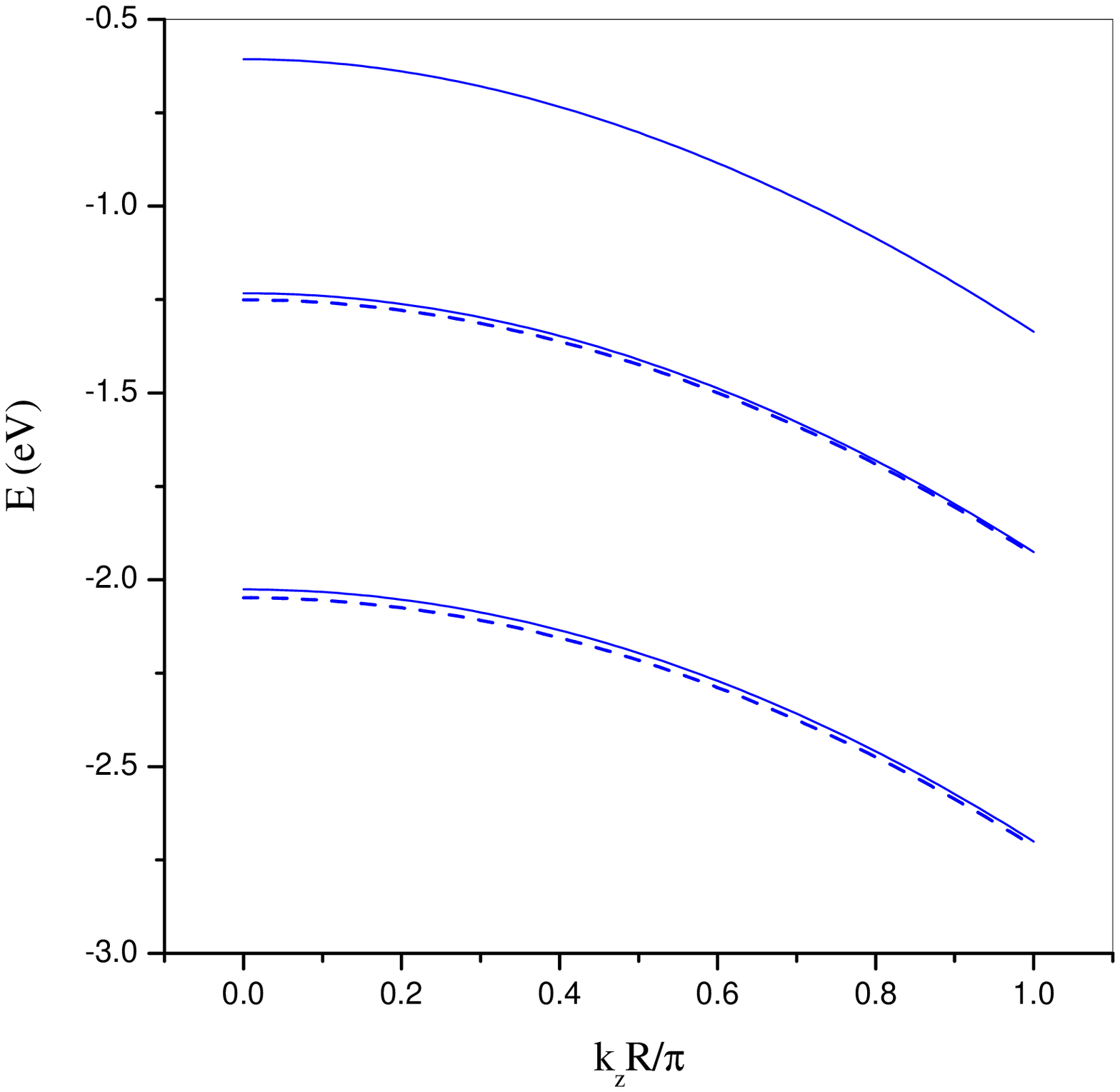}
\caption{(Color online) Energy dispersion of electron subbands
within the valence band of a PbSe
cylindrical quantum wire of radius $R=20$~\AA. By dashed lines are shown subbands
with the quantum number $n_v=2$.}
\end{figure}

The condition that this system of algebraic equations has a non-trivial
solution yields
\begin{equation}
\label{kappa}
\kappa^2-k_z^2=\Xi-\Lambda \,.
\end{equation}
Equations~(\ref{sys3}),~(\ref{sys4}) allow one to express the coefficients
$F$ and $G$ in terms of $C$ and $D$:
\[
F=\frac{P(-k_z \, C+i \, \kappa \, D)}{\alpha_v \, \kappa^2 - \alpha_v \, k_z^2-E-E_g/2} \,,
\]
\[
G=\frac{P(i \, \kappa \, C+ k_z \, D)}{\alpha_v \, \kappa^2 - \alpha_v \, k_z^2-E-E_g/2} \,.
\]
By setting zero either $C$ or $D$ we arrive to the following solutions
of Eqs.~(\ref{dimmock})
\begin{equation}
\label{sol3u}
\hat{u}_{M}^{(3)} (\rho,\varphi,z)=C \, e^{i \, k_z \, z}
\left[
\matrix{
e^{i \,(M-1/2) \, \varphi} \, I_{M-1/2}(\kappa \rho) \cr
0 \cr
}
\right] \,,
\end{equation}
\begin{equation}
\label{sol3v}
\hat{v}_{M}^{(3)} (\rho,\varphi,z)=\frac{i \, P \, C \, e^{i \, k_z \, z}}
{\alpha_v \, \kappa^2 - \alpha_v \, k_z^2-E-E_g/2}
\,
\left[
\matrix{
i \, k_z \, e^{i \, (M-1/2) \, \varphi} \, I_{M-1/2}(\kappa \rho) \cr
\kappa \, e^{i \, (M+1/2) \, \varphi} \, I_{M+1/2}(\kappa \rho) \cr
}
\right] \,,
\end{equation}
\begin{equation}
\label{sol4u}
\hat{u}_{M}^{(4)} (\rho,\varphi,z)=D \, e^{i \, k_z \, z}
\left[
\matrix{
0 \cr
e^{i \, (M+1/2) \, \varphi} \, I_{M+1/2}(\kappa \rho) \cr
}
\right] \,,
\end{equation}
\begin{equation}
\label{sol4v}
\hat{v}_{M}^{(4)} (\rho,\varphi,z)=\frac{i \, P \, D \, e^{i \, k_z \, z}}
{\alpha_v \, \kappa^2 - \alpha_v \, k_z^2-E-E_g/2}
\,
\left[
\matrix{
\kappa \, e^{i \, (M-1/2) \, \varphi} \, I_{M-1/2}(\kappa \rho) \cr
-i \, k_z \, e^{i \, (M+1/2) \, \varphi} \, I_{M+1/2}(\kappa \rho) \cr
}
\right] \,.
\end{equation}

The four solutions we constructed are not mutually orthogonal but they are
linearly independent. If one requires that their linear combination
(with the coefficients $A$, $B$, $C$, and $D$) vanishes
at $\rho=R$ then one will obtain a system of four homogeneous algebraic
equations on these coefficients. The condition that this system has a
non-trivial solution will lead to the dispersion equation determining the
allowed energy values of electrons confined in a cylindrical
quantum wire. The positive (negative) values of energy describe the
states in the conduction (valence) band. The dispersion equation takes
the form
\[
\alpha_v^2 \, k_z^2 \, (k^2 + \kappa^2)^2 \, J_{M-1/2}(kR) \,
J_{M+1/2}(kR) \, I_{M-1/2}(\kappa R) \, I_{M+1/2}(\kappa R)
\]
\[
+\left[ k \, (\alpha_v \, \kappa^2 - \alpha_v \, k_z^2-E-E_g/2) \,
J_{M-1/2}(kR) \, I_{M+1/2}(\kappa R)  \right.
\]
\begin{equation}
\label{disp}
\left. + \kappa \,
(\alpha_v \, k^2 + \alpha_v \, k_z^2+E+E_g/2) \,
J_{M+1/2}(kR) \, I_{M-1/2}(\kappa R) \right]
\end{equation}
\[
\times \left[ k \, (\alpha_v \, \kappa^2 - \alpha_v \, k_z^2-E-E_g/2) \,
J_{M+1/2}(kR) \, I_{M-1/2}(\kappa R) \right.
\]
\[
\left. - \kappa \,
(\alpha_v \, k^2 + \alpha_v \, k_z^2+E+E_g/2) \,
J_{M-1/2}(kR) \, I_{M+1/2}(\kappa R) \right]=0 \,.
\]

In order to better understand the structure of this equation let
us first consider the case of $k_z=0$. In this limit each of the
spinors $\hat{u}_M^{(i)}$, $\hat{v}_M^{(i)}$ ($i=1,2,3,4$) has only
one non-zero component. Therefore, for every $i$ the corresponding
bispinor solution has two non-zero components. Taking into account the parity
of the Bloch functions $|L_6^{\mp} \rangle$,
the solutions with $i=1$ and $i=3$ have in the
limit $k_z=0$ the parity $(-1)^{M+1/2}$, while the solutions with $i=2$ and
$i=4$ have the parity $(-1)^{M-1/2}$. The first term of Eq.~(\ref{disp})
vanishes in the limit $k_z=0$, and Eq.~(\ref{disp}) reduces to a product
of the two square brackets. The first (second) square bracket is responsible
for the
solutions with the parity $(-1)^{M \mp 1/2}$. This situation is analogous
to the case of spherical quantum dots~\cite{kang,myPbSe}.

When $k_z \neq 0$ then each bispinor solution we constructed has three
non-zero components. In this case the first term in Eq.~(\ref{disp})
does not vanish and factorization of the dispersion equation becomes
impossible. This observation is in contradiction with the conclusions
of Ref.~\cite{rupasov}.

Because of the Kramers degeneracy Eq.~(\ref{disp}) must be invariant
under the change of $M$ to $-M$. This invariance is guaranteed by
the properties of the Bessel functions: $J_{-n}(x)=(-1)^n \, J_n(x)$;
$I_{-n}(x)=I_n(x)$.

Let us illustrate our results by numerical calculations.
In Fig.~1 are shown energies of electron states in a PbSe
cylindrical quantum wire at $k_z=0$
as a function of the wire radius. Only dependences for the five lowest
subbands in the conduction band and five uppermost subbands in the
valence band are shown.
All the material parameters are taken
from Ref.~\cite{kang}. The electron states in the conduction (valence) band
are characterized by the
projection, $M$ of the total angular momentum onto the wire axis
and by the main quantum number $n_c$ ($n_v$).

In Fig.~2 (Fig.~3) is shown energy dispersion of electron subbands
within the conduction (valence) band of a PbSe
cylindrical quantum wire of radius $R=20$~\AA. The solid lines refer to
the subbands with the main quantum number $n_{c(v)}=1$ while the dashed
lines refer to
the subbands with the main quantum number $n_{c(v)}=2$. The dependences
shown in Figs.~2,~3 are nearly parabolic.

To summarize, we have studied electronic structure of cylindrical lead salt
quantum wires within the multi-band envelope-function formalism. We have
found the energy dispersion for subbands within the conduction and valence
bands of the quantum wire. When applied to real structures, the parameters
of the model may be modified for each particular $L$-valley to reflect its
orientation with respect to the quantum wire growth direction and taking
into account the valley anisotropy.

The author wishes to thank A.N.~Poddubny for useful discussions.
This work was supported in part by the NSF under grant No.~HRD-0833178
and in part by the Russian Foundation for Basic Research.


\begin{thebibliography}{99}
\bibitem{fn1} We use the system of units where $\hbar=m_0=e=1$ ($m_0$ and
$e$ are the mass and the charge of a free electron).
\end{thebibliography}
\end{document}